\documentclass[useAMS,usenatbib,onecolumn]{mn2e}

\usepackage{graphicx}
\usepackage{dcolumn}
\usepackage{bm}
\usepackage{natbib}
\bibpunct{(}{)}{,}{a}{}{,}
\begin{document}


\title{Generalised Inverse-Cowling Approximation for Polar $w$-mode Oscillations of
Neutron Stars}

\author[J.~Wu and P.T.~Leung]{J.~Wu\thanks{Present address: Department of Physics and Astronomy,
University of Pittsburgh, Pittsburgh, Pennsylvania 15260, USA.}
and P.T.~Leung\thanks{Correspondence author (email:
ptleung@phy.cuhk.edu.hk).}\\
Physics Department and Institute of Theoretical Physics, The
Chinese University of Hong Kong, Shatin, Hong Kong SAR, China.}

\date{\today}
\pagerange{\pageref{firstpage}--\pageref{lastpage}} \pubyear{2007}

\label{firstpage}
\def\tomega{ \tilde{\omega} }
\def\tOmega{ \tilde{\Omega} }
\def\tr{ \tilde{r} }
\def\tx{ \tilde{r}_* }
\def\tV{ \tilde{V} }
\def\tpsi{ \tilde{\psi} }
\def\trho{ \tilde{\rho} }
\def\tP{ \tilde{P} }
\def\tR{ \tilde{R} }
\def\tX{ \tilde{R}_* }
\def\tm{ \tilde{m} }
\def\tphi{ \tilde{\nu} }
\def\tlam{ \tilde{\lambda}}
\def\tepsilon { \tilde{\epsilon}}
\def\tHO{ \tilde{H}_0}
\def\tHI{ \tilde{H}_1}
\def\tK{ \tilde{K}}
\def\tW{ \tilde{W}}
\def\tX{ \tilde{X}}
\def\tV{ \tilde{V}}
\def\tgamma{ \tilde{\gamma}}
\def\ta{ \tilde{a}}
\def\tb{ \tilde{b}}
\def\tg{ \tilde{g}}
\def\th{ \tilde{h}}
\def\tk{ \tilde{k}}
\def\a{ {\rm a}}
\def\c{ {\rm c}}
\def\e{ {\rm e}}
\def\p{ {\rm p}}
\def\f{ {\rm f}}
\def\d{ {\rm d}}
\def\i{ {\rm i}}
\def\r{ {\rm r}}
\maketitle
\begin{abstract}
Adopting the Lindblom-Detweiler formalism for polar oscillations
of neutron stars, we study the $w$-mode oscillation and find that
the Lagrangian change in pressure, measured by the physical
quantity $X$, is negligibly small. Based on this observation, we
develop the generalised inverse-Cowling approximation (GICA) with
the approximation $X=X'=0$, where $X'$ is the derivative of $X$
with respect to the circumferential radius, for $w$-mode
oscillations of neutron stars. Under GICA, $w$-mode oscillations
are described by a second-order differential system, which can
yield accurate frequencies and damping rates of quasi-normal
modes.
\end{abstract}

\begin{keywords}
gravitational waves - quasi-normal modes of compact stars -
 - stars: neutron - stars: oscillations - relativity
\end{keywords}
\section{Introduction}
The search for gravitational waves emitted from stellar collapse
has been the major goal of many relativists and astrophysicists
since the early seventies of last century. At the beginning of a
new millenium, several ground-based interferometric detectors
(e.g. LIGO, VIRGO, GEO 600, TAMA 300) are currently endeavouring
to detect gravitational waves from extraterrestrial objects
\citep[see, e.g.,][and references therein]{Hughes_03,Grishchuk}.
With the upgrade of these detectors and the availability of a
space-based interferometric detector, LISA, which is expected to
be launched at the beginning of the next decade, it is widely
believed that gravitational wave detection and hence astronomy in
the gravitational wave window will become achievable in the near
future \citep{Hughes_03,Grishchuk}. In fact, there have been four
science runs on LIGO and GEO 600 since 2002. Despite that so far
no positive evidence for the existence of gravitational waves has
been found, relevant observational data were used to place limits
on the ellipticity and the gravitational wave strain of a number
of known pulsars \citep{LIGO_05,LIGO_07}.

One major kind of gravitational wave emitting astrophysical
processes is certainly stellar gravitational collapse, often
resulting in remnants of neutron stars or black holes, which are
also possible gravitational wave emitters \citep[for a review on
gravitational waves emitted in gravitational collapse see,
e.g.,][and references therein]{Fryer_rev}. For example,
gravitational waves emitted in the formation of neutron stars in
our galaxy via stellar core collapses have been shown to be
detectable for sufficiently fierce asymmetric collapse
\citep{Fryer:2001zw,Lindblom:1998wf}. Since such gravitational
waves are likely to carry the information about the internal
structure of the neutron star involved, such as its mass, radius,
density distribution and in turn its equation of state (EOS), the
characteristics of gravitational wave signals emitted from
undulating neutron stars have become a subject of much study
\citep[see,
e.g.,][]{Andersson1996,Andersson1998,Ferrari,Kokkotas_2001,Ferrari_prd}.

The spectrum of gravitational waves from compact stellar objects
are commonly described in terms of quasi-normal modes (QNMs). Each
of these QNMs is characterized by a complex eigenfrequency
$\omega=\omega_\r+\i\omega_\i$ and has a time dependence
$\exp(\i\omega t)$
\citep{Press_1971,Leaver_1986,Ching,Kokkotas_rev}. The QNM
frequencies of gravitational waves emitted from a neutron star are
generally model-sensitive and are expected to depend significantly
on the EOS used in the stellar model. In spite of this fact, some
universal behaviours in the frequency $\omega_\r$ and the damping
time $\tau\equiv 1/\omega_\i$ of the $f$-mode and the $w$-mode
oscillations of non-rotating neutron stars can still be identified
\citep{Andersson1998,Ferrari}, which have recently been explained
by \citet{Tsui05:1}. Based on such universal behaviours, it has
been shown that the radius and the mass of a neutron star can be
inferred from the pulsation frequencies of its fundamental fluid
$f$-mode and the first $w$-mode \citep{Andersson1998,Ferrari}.
Most interestingly, \citet{Tsui05:3} and \citet{Tsui:prd} showed
that the EOS of a neutron star can be inferred from the
eigenfrequencies of a few (say, e.g., 3) $w$-mode QNMs. These
discoveries all underscore the physical significance of QNMs of
neutron stars. On the other hand, it is important to note that
detection of gravitation waves from neutron stars with sufficient
accuracy to infer their internal structure is a mission impossible
in the short (or even medium) term. As estimated by
\citet{Comer_1} and \citet{Tsui:prd}, realisation of such schemes
has to await the availability of more advanced gravitational-wave
observatory like EURO.

Non-radial oscillations of relativistic neutron stars, first
studied by \citet{Thorne67}, can be analysed by decomposing the
perturbed Einstein equations into spherical harmonics. Such
oscillations are categorized into axial (toroidal) and polar
(spheroidal) modes according to the parity of the harmonics. In
axial-mode oscillations, which can be described by a single
second-order wave equation \citep{Chandrasekhar1}, fluid elements
of a star are merely spectators in the sense that they are not
affected by gravitational waves generated \citep{Thorne67}. On the
other hand, in polar-mode oscillations, fluid motion and
gravitational wave are coupled together. Due to the complicated
interplay between the matter and the metric, polar-mode
oscillations are usually governed by a system of ordinary
differential equations (ODEs) involving several physical
variables. In the pioneering work of \citet{Thorne67}, polar
oscillations are described by a fifth-order system of ODEs. Later,
\citet{Lindblom_1983} explicitly reduced the equation set into a
fourth-order one, hereafter referred to as the LD formalism, where
the oscillation of a star is described by four independent
physical quantities $H_1$, $K$, $W$, and $X$. While $H_1$ and $K$
are measures of the metric, $W$ and $X$ are respectively
proportional to the Lagrangian displacement in the radial
direction and the pressure change of the fluid constituting the
star (see Sect. 2 for more details). Besides the scheme proposed
by \citet{Lindblom_1983}, there are other schemes describing polar
oscillations with two coupled second-order ODEs
\citep{Price,Price2,PhysRevD.46.4289,AAKS98}. They are in fact
equivalent to fourth-order systems.

 As in the Newtonian theory of stellar pulsations, polar
oscillations can be classified into the fundamental ($f$) mode,
the pressure ($p$) mode, and the gravity ($g$) mode. In addition
to these, there is an extra kind of mode in the relativistic
theory, namely the spacetime ($w$) mode, representing mainly
perturbations in the metric
\citep{Kojima_1988,Kokkotas_Schutz,Andersson1995,Andersson1996}.

Soon after the discovery of the polar $w$-mode, it has been
conjectured that the fluid in a relativistic star will not be
affected much by $w$-mode pulsations
\citep{Kokkotas_Schutz,Andersson1995}. Aiming at understanding
qualitatively the physical nature of polar $w$-mode pulsations,
\citet{in_Cowling} suggested the Inverse-Cowling Approximation
(ICA) for polar $w$-mode oscillations. In ICA the fluid motion is
completely neglected from the outset (details of ICA will be
discussed in Sect. 2 of the present paper) and the metric
variables are governed by a second-order ODE system. The
approximation can successfully reproduce qualitative features of
polar $w$-mode oscillations and clarify the nature of the
oscillations \citep{Andersson:1995ez}. However, as shown in
Fig.~1, it fails to reproduce accurate values of the QNM
frequencies. This is a sort of expected because the main concern
of ICA was qualitative discussion on polar $w$-mode oscillations.

The major objective of the present paper is to establish a
quantitatively correct approximation for the polar $w$-mode,
referred to as Generalised Inverse-Cowling Approximation (GICA) in
the present paper, which is capable of locating polar $w$-mode
with high accuracy (see Fig.~1). In our analysis we find that the
fluid motion of a star is in fact non-negligible. However, $X$ is
so small that  the eigenfrequency of polar $w$-mode can be
accurately determined with the approximation $X=X'=0$ (hereafter a
prime indicates differentiation with respect to circumferential
radius $r$) by solving two coupled first-order ODEs. Thus, under
GICA polar $w$-mode oscillations are describable with one single
second-order ODE system.

 The organization of our paper is as follows. In Sect.~2, we
 briefly review the LD formalism. The ICA developed by
 \citet{in_Cowling} is discussed in Sect.~3 and we propose GICA
 in Sect.~4.
We then conclude our paper in Sect.~5 with a short conclusion and
discussion. Unless otherwise stated, geometrized units in which
$G=c=1$ are adopted in the following discussion.

\section{LD Formalism}
In spherical coordinates, $(t,r,\theta,\varphi)$, the geometry of
spacetime around a non-rotating unperturbed neutron star is given
by the line element:
\begin{equation}
\d s^2=-\e^{\nu(r)}\d t^2+\e^{\lambda(r)}\d
r^2+r^2(\d\theta^2+\sin^2\theta \d\varphi^2).
\end{equation}
The metric coefficient $\e^{\lambda(r)}$ is determined by the mass
distribution function $m(r)$, the mass inside circumferential
radius $r$,
\begin{eqnarray}
    \e^{-\lambda(r)}&=&1-\frac{2m(r)}{r}. \label{g1}
\end{eqnarray}
The other metric coefficient $\e^{\nu(r)}$ and the mass
distribution function $m(r)$ can be obtained from the solution to
the Tolman-Oppenheimer-Volkoff (TOV) equations
\citep{Tolman:1939jz,Oppenheimer:1939ne}:
\begin{eqnarray}
\frac{\d\nu}{\d r} &=& \frac{2m+8\pi r^3 p}{r(r-2m)} ,\label{g2}\\
\frac{\d m}{\d r} &=& 4\pi r^2\rho, \label{TOV1}\\
\frac{\d p}{\d r} &=& -\frac{1}{2}(\rho+p)\frac{\d\nu}{\d
 r},\label{TOV2}
\end{eqnarray}
where $\rho$ and $p$ are the mass density and pressure,
respectively.

On the other hand, in the LD formalism for a pulsating
relativistic star with radius $R$ \citep{Lindblom_1983}, the
perturbation in the metric, $h_{\mu\nu}$, can be expressed in
terms of three physical quantities $H_0$, $H_1$ and $K$ and
displayed in a matrix form as follows:
\begin{equation}\label{metric}
h_{\mu\nu} = \left( \begin{array}{cccc}
-\e^{\nu}\mu^lH_0&-\i\omega \mu^lH_1&0&0\\-\i\omega
\mu^lH_1&-\e^{\lambda}\mu^lH_0&0&0\\0&0&-\mu^lr^{2}K&0\\
0&0&0&-\mu^lr^{2}\sin^2\theta K \end{array} \right)\e^{\i\omega
t}Y_{lm}(\theta,\varphi),
\end{equation}
where $\omega$ is eigenfrequency of the polar mode oscillation,
$l$ is the angular momentum quantum number, and
\begin{eqnarray}
\mu = \left\{ \begin{array}{lr} \displaystyle{\frac{r}{R}} &
\hspace{0.5cm}r<R; \\ 1 & \hspace{0.5cm}r \ge R. \end{array}
\right.
\end{eqnarray}
 Correspondingly, the Lagrangian displacements
of the fluid are expressed by:
\begin{eqnarray}
\xi^r & = &
\mu^lr^{-1}\e^{-\lambda/2}WY_{lm}(\theta,\varphi)\e^{\i\omega
t},\\
\xi^{\theta} & = &
-\mu^lr^{-2}V\partial_{\theta}Y_{lm}(\theta,\varphi)\e^{\i\omega
t},\\
\xi^{\varphi} & = &
-\mu^l(r\sin\theta)^{-2}V\partial_{\varphi}Y_{lm}(\theta,\varphi)\e^{\i\omega
t}.
\end{eqnarray}
In other words, $W$ and $V$ together measure the spatial
displacements of the fluid.

Putting all these perturbations into the Einstein equations,
\citet{Lindblom_1983} showed that these physical variables are
related by four first-order ODEs:
\begin{eqnarray}
H'_1 & = & -\frac{1}{r}\left[ l+1+\frac{2m\e^{\lambda}}{r}+4\pi
r^2\e^{\lambda}(p-\rho) \right]H_1
        +\frac{1}{r}\e^{\lambda}[H_0+K-16\pi(\rho+p)V],\label{Lindblom-1} \\
K' & = & \frac{1}{r}H_0+\frac{1}{2r}l(l+1)H_1-\left[ \frac{l+1}{r}-\frac{
\nu'}{2}\right]K-\frac{8\pi(\rho+p)\e^{\lambda/2}}{r}W, \label{Lindblom-2} \\
W' & = & -\frac{l+1}{r}W+r\e^{\lambda/2}\left[ \frac{\e^{-\nu/2}}{\gamma p}X-\frac{l(l+1)}{r^2}V+\frac{1}{2}H_0+K\right], \label{Lindblom-3} \\
X' & = &
-\frac{l}{r}X+(\rho+p)\e^{\nu/2}\left\{\frac{1}{2}\left(\frac{1}{r}-\frac{\nu'}{2}\right)H_0+
\frac{1}{2}\left[
r\omega^2\e^{-\nu}+\frac{1}{2}\frac{l(l+1)}{r}\right]H_1 \right.
 +\frac{1}{2}\left(\frac{3}{2}\nu'-\frac{1}{r}\right)K-\frac{1}{2}\frac{l(l+1)}{r^2}\nu'V \nonumber \\
& & \left.-\frac{1}{r}\left[  4\pi(\rho+p)\e^{\lambda/2}
+\omega^2\e^{\lambda/2-\nu} -\frac{1}{2}r^2\frac{\d}{\d r}
\left(\frac{\e^{-\lambda/2}\nu'}{r^2}\right)\right]W  \right\}, \label{Lindblom-4} \\
\nonumber
\end{eqnarray}
where $\rho(r)$, $p(r)$ are respectively the density and the
pressure at a radius $r$, $\gamma$ is the adiabatic index, defined
by
\begin{equation}
\gamma = \frac{\rho+p}{p}\frac{dp}{d\rho},
\end{equation}
and
\begin{equation}
X =
\omega^2(\rho+p)\e^{-\nu/2}V-\frac{1}{r}p'\e^{(\nu-\lambda)/2}W
+\frac{1}{2}(\rho+p)\e^{\nu/2}H_0.\label{Lindblom-5}
\end{equation}
The physical significance of $X$ is that it is a measure of
$\Delta p$, the Lagrangian change in the fluid pressure. In fact,
it can be readily shown  from the theory developed by
\citet{Thorne67} and \citet{Lindblom_1983} that:
\begin{equation}\label{pressure}
\Delta p=-\e^{-\nu/2}\mu^l X Y_{lm}(\theta,\varphi)\e^{\i\omega
t}.
\end{equation}
 Besides, the metric perturbation
$H_0$ is expressed in terms of $H_1$, $K$ and $X$:
\begin{eqnarray}
\left[ 3m+\frac{1}{2}(l+2)(l-1)r+4\pi r^3p \right]H_0
&  =& 8\pi r^3\e^{-\nu/2}X-\left[ \frac{1}{2}l(l+1)(m+4\pi r^3p)-\omega^2r^3\e^{-(\nu+\lambda)}\right]H_1 \nonumber \\
& & + \left[ \frac{1}{2}(l+2)(l-1)r-\omega^2r^3\e^{-\nu}-
\frac{1}{r}\e^{\lambda}(m+4\pi r^3p)(3m-r+4\pi r^3p)
\right]K.\label{Lindblom-6}
\end{eqnarray}
Eqs. (\ref{Lindblom-1}) - (\ref{Lindblom-4}), supplemented by the
algebraic equations (\ref{Lindblom-5}) and (\ref{Lindblom-6}),
then form a non-singular fourth-order system of ODEs with $\{H_1,
K, W, X\}$ being the four independent variables \citep{Detweiler}.
These four ODEs plus (i) the regularity condition at $r=0$, (ii)
the condition $X=0$ at $r=R$, and (iii) the outgoing boundary
condition at $r=\infty$ completely determine the eigenfrequency of
polar mode pulsations \citep{Lindblom_1983}.
\section{Inverse-Cowling  Approximation}
Working in LD formalism and motivated by the conjecture that the
fluid in a relativistic star is hardly excited by $w$-mode
pulsations \citep{Kokkotas_Schutz,Andersson1995},
\citet{in_Cowling} attempted to study the physical nature of polar
$w$-mode oscillations by proposing the ICA with $V=W=0$. Under
such assumption, the motion of the fluid is completely ignored. To
gauge the accuracy of the assumption $V=W=0$, we show in Fig.~1
the $l=2$ (always assumed for all numerical results in this paper)
polar $w$-mode QNM frequencies of (a) a polytropic star with $p
=\kappa \rho^2$ and $\kappa=100$, which was used in the paper by
\citet{in_Cowling}, and (b) an APR1 star \citep{APR}, which
typifies other realistic neutron stars, and compare the numerical
results obtained respectively from exact LD formalism, ICA with
$V=W=0$ and GICA (to be developed in Sect.~4 of the present
paper). It is obvious that in both cases the eigenfrequency
obtained from the assumption $V=W=0$ deviates markedly from the
exact one and the agreement between the approximate and exact
values is only qualitative. As noted by \citet{in_Cowling}, the
imaginary parts of QNM frequencies obtained from ICA are typically
20-30 percent smaller than the exact values and the relative error
in the frequency spacing is about 10 percent. This is not really
surprising because the original motivation of ICA was merely to
verify the spacetime nature of polar $w$-mode oscillations.

 The inadequacy of the approximation $V=W=0$
can be clearly shown by comparing the contributions of various
terms in (\ref{Lindblom-5}), whose physical meaning will be
discussed at the end of this paper. In Fig.~2, the absolute values
of
 $\omega^2(\rho+p)\e^{-\nu/2}V$, $p'\e^{(\nu-\lambda)/2}W/r$,
$(\rho+p)\e^{\nu/2}H_0/2$ and $X$ are compared for the a typical
polar $w$-mode of the APR1 neutron star considered in Fig.~1b. We
note the term $\omega^2(\rho+p)\e^{-\nu/2}V$, a measure of the
fluid motion in the tangential direction, is comparable to the
term $(\rho+p)\e^{\nu/2}H_0/2$, a perturbation in the metric.
Besides, the term $p'\e^{(\nu-\lambda)/2}W/r$, measuring the
motion in the radial direction, is not as small as expected.
Judging from the results shown in Figs.~1 and 2, we deem that the
approximation $V=W=0$  is unjustified. Instead, as suggested by
the comparison shown in Fig.~2, the approximation $X=X'=0$ seems
to be more appropriate. Similar behaviour also prevails in stars
constructed with other realistic EOSs. We note that this point has
been discussed and exploited by \citet{Tsui05:1} to explain the
universality observed in the polar $w$-mode.  In the following
section we will base on GICA with $X=X'=0$ to establish a
second-order ODE system that is able to locate $w$-mode QNMs
accurately.

\section{Generalised Inverse-Cowling  Approximation}
Under GICA where $X=X'=0$, (\ref{Lindblom-4}) and
(\ref{Lindblom-5}) yield respectively
\begin{eqnarray}
0 & = & \frac{1}{2}(\frac{1}{r}-\frac{\nu'}{2})\,H_0+ \frac{1}{2}\left[ r\omega^2\e^{-\nu}+\frac{1}{2}\frac{l(l+1)}{r}\right]\,H_1 +\frac{1}{2}(\frac{3}{2}\nu'-\frac{1}{r})\,K-\frac{1}{2}\frac{l(l+1)}{r^2}\nu'\,V \nonumber \\
& &
-\frac{1}{r}\left[4\pi(\rho+p)\e^{\lambda/2}+\omega^2\e^{\lambda/2-\nu}
-\frac{1}{2}r^2\frac{\d}{\d
r}\left(\frac{\e^{-\lambda/2}\nu'}{r^2}\right)\right]\,W;
\end{eqnarray}
and
\begin{equation}
V = \frac{-\e^{\nu}}{2{\omega}^2}\,H_0 -
\frac{\e^{\nu-\lambda/2}\left( m + 4p\pi r^3\right)}{r^2\left(-2m
+ r\right){\omega}^2}\,W. \label{V-exp}
\end{equation}
Solving these two equations, we can express $W$ in terms of the
metric coefficients $H_0$, $H_1$ and $K$:
\begin{eqnarray}
W & = & \frac{1}{D}\left\{\e^{\lambda/2}\left[\frac{-5m+r-12p\pi r^3}{2(r-2m)}\,K+\frac{1}{4}\left[l(l+1)+2\e^{-\nu}r^2\omega^2\right]\,H_1 \right.\right. \nonumber \\
& & \left.\left. +\frac{\left[-\e^{\nu}l(l+1)(m+4\pi
pr^3)\right]+r^2\omega^2(3m-r+4\pi pr^3)}{2(r-2m)r^2\omega^2}\,H_0
\right]\right\} \,,\label{W-exp}
\end{eqnarray}
where
\begin{equation}\label{}
D=-4\pi \e^{\lambda}(p+\rho)+\frac{7m^2-4\pi r^4(p-\rho+4\pi
p^2r^2)-4m(r+2\pi\rho r^3)}{r^2(r-2m)^2}\,.
\end{equation}
Moreover, following directly from (\ref{Lindblom-6}) and $X=0$,
$H_0$ becomes:
\begin{eqnarray}
H_0 & = & \frac{1}{{3m + \frac{1}{2}(l+2)(l-1)r + 4p\pi
r^3}}\left\{
 \left[\frac{-l(l+1)}{2}(m + 4\pi p r^3) + \e^{-(\lambda+\nu)}r^3\omega^2 \right]\,H_1 \right. \nonumber \\
& & \left. + \left[\frac{1}{2}\left( l-1  \right)\left( l+2  \right)r - \frac{\e^{\lambda }}{r}\left( m + 4 p\pi r^3\right)\left( 3m - r + 4 p\pi r^3 \right) - r^3{\omega }^2\e^{-\nu}\right]\,K\right\}, \nonumber \\
\label{H0-exp}
\end{eqnarray}
As a result, $V$, $W$ and $H_0$ are all expressible in terms of
$H_1$ and $K$, which are governed by two first-order ODEs
(\ref{Lindblom-1}) and (\ref{Lindblom-2}).

To solve this second-order ODE system, we first discuss the
boundary conditions at the center $r=0$. Straightforward expansion
about the center shows that Eqs.~(\ref{Lindblom-1}) and
(\ref{Lindblom-2}) lead to:
\begin{eqnarray}
0 & = & -\frac{l+1}{r}\,H_1(0)+\frac{1}{r}\left[H_0(0)+K(0)-16\pi(p_0+\rho_0)\,V(0)\right], \label{H1-eqn-center} \\
0 & = &
\frac{1}{r}\,H_0(0)+\frac{1}{2r}l(l+1)\,H_1(0)-\frac{l+1}{r}\,K(0)-\frac{8\pi(p_0+\rho_0)}{r}\,W(0),
\label{K-eqn-center}
\end{eqnarray}
Comparing (\ref{H1-eqn-center}) and (\ref{K-eqn-center}), we see
that:
\begin{equation}
W(0) = -l\,V(0). \label{W-center}
\end{equation}
In addition, (\ref{H0-exp}) gives
\begin{equation}
K(0) = H_0(0), \label{H0-center}
\end{equation}
at the center. Putting (\ref{W-center}) and (\ref{H0-center}) back
into (\ref{V-exp}), we have:
\begin{equation}
K(0) = H_0(0) =
\left[-\frac{8\pi}{3}(3\,p+\rho)+\frac{2\,\omega^2\e^{-\nu}}{l}\right]\,W(0).
\end{equation}
Finally, from (\ref{H1-eqn-center}) or (\ref{K-eqn-center}),
\begin{equation}
H_1(0) =
\frac{1}{l(l+1)}\left[2l\,K(0)+16\pi(p+\rho)\,W(0)\right].
\end{equation}
Therefore, for a  given $W(0)$, which merely provides a scale of
the solution, the values of $H_1(0)$, $K(0)$, $H_0(0)$ and $V(0)$
are all determined. Compared with the boundary conditions at the
center, the boundary condition at the stellar surface $r=R$ is
much simpler. There are two independent solutions near the surface
with arbitrarily assigned values of $H_1(R)$ and $K(R)$.
Integrating (\ref{Lindblom-1}) and (\ref{Lindblom-2}) outward from
$r=0$ and inward from $r=R$, and matching the values of $H_1$, $K$
at some intermediate value of $r$, say $r=R/2$, can numerically
determine the solution of the second-order system up to a
multiplicative constant.

The QNM frequencies obtained from GICA are shown in Fig.~1 and
Table~1, which agree nicely with the exact values. In particular,
both the percentage errors in the frequency and the damping rate,
$E_\r$ and $E_\i$, decrease with increasing frequency, reflecting
the decrease in the Lagrangian change in pressure in
high-frequency polar $w$-modes. Meanwhile, it is also worthy to
note that even the $w_{\rm II}$ mode, characterized by large
damping rates, can be reproduced accurately. Hence, GICA proposed
here, which assumes $X=0$, clearly outperforms the ICA with the
approximation $V=W=0$.

\section{Conclusion and Discussion}
We have developed in the present paper an accurate scheme, GICA,
to describe $w$-mode oscillations of neutron stars. Instead of
completely ignoring the displacement of the fluid element of a
star under consideration, we have shown that the term $X$, which
is proportional to the Lagrangian change in the pressure, is
negligible. Under GICA, $w$-mode oscillations of neutron stars are
governed by two first-order ODE in the metric coefficients $H_1$
and $K$. The displacements of matter, measured by $V$ and $W$, are
properly considered and found to be expressible in terms of $H_1$
and $K$. Numerical results obtained directly from the
approximation $X=X'=0$ clearly demonstrate the validity and
accuracy of GICA (see Fig.~1 and Table~1).

In addition to yielding reliable numerical values of QNM
frequencies for $w$-mode oscillations, GICA proposed here also
provides a proper physical picture for such oscillations. A star
undergoing polar $w$-mode oscillations has non-negligible
displacements in the tangential and radial directions. However, as
shown in Fig.~2, such displacements are driven by the perturbed
spacetime, instead of the excess pressure due to the motion of
matter.

It is well known that axial $w$-mode oscillations of compact
stars, which are merely oscillations in the metric variables, are
governed by a second-order Regge-Wheeler wave equation
\citep{Chandrasekhar1}. Our finding that polar $w$-mode
oscillations are adequately described by one single second-order
ODE system thus puts polar and axial $w$-modes on an equal footing
in the sense that both of them are governed by second-order ODE
systems. As shown in the review by \citet{Kokkotas_rev}, QNMs of
these two kinds of $w$-mode lie asymptotically on a common smooth
curve in the complex frequency plane. In addition,
\citet{Kojima_95} and \citet{Andersson:1995ez} also noted that for
incompressible ultra compact stars  the oscillation frequencies of
these two $w$-modes agree to within one percent. GICA proposed
here may be the first step in the long process of seeking
quantitative interpretation  of such interesting asymptotic
behaviour and the similarity between two kinds of $w$-mode.

We have adopted LD formalism to describe polar oscillations in the
present paper. In fact, there are other alternative formalisms for
polar oscillations of relativistic stars \citep[see,
e.g.,][]{Price,PhysRevD.46.4289,AAKS98,2004PhRvD..69l4028N}. In
particular, in the formalism proposed by \citet{AAKS98}, and later
used by \citet{2004PhRvD..69l4028N} to study accretion-driven
gravitational waves from compact stars, such oscillations are
described by two coupled second-order wave equations in the metric
variables plus a Hamiltonian constraint yielding the Eulerian
change in the fluid pressure. The approach is also equivalent to a
fourth-order differential equation system with the tortoise
ordinate being the independent variable. The formalism is
evidently more amenable to decoupling metric and matter variables.
It is worthy to develop quantitatively accurate approximation
scheme for polar $w$-mode oscillations based on such formalism.
Currently we are working along this direction.

Lastly, it is worthy to note the physical meaning of
(\ref{Lindblom-5}), which can be rewritten as follows:
\begin{equation}
\mu^l\omega^2(\rho+p)V Y_{lm}(\theta,\varphi)\e^{\i\omega t} =
-\e^{\nu}\delta p -\frac{\mu^l (\rho+p)\e^{\nu} H_0}{2}
Y_{lm}(\theta,\varphi)\e^{\i\omega t},\label{Lindblom-5-New}
\end{equation}
where
\begin{equation}\label{}
\delta p \equiv \Delta p-p'\xi^{r}= \Delta p- \frac{\mu^l p'
\e^{-\lambda/2}W}{r} Y_{lm}(\theta,\varphi)\e^{\i\omega t}
\end{equation}
 is the Eulerian change in
pressure. Each term in (\ref{Lindblom-5-New}) has an obvious
analogue in Newtonian theory of stellar pulsation \citep[see, e.g.
Eq. (17.21b) in][]{Cox}. The term
$\mu^l\omega^2(\rho+p)VY_{lm}(\theta,\varphi)\e^{\i\omega t}$ in
the LHS is a measure of ${\rm inertia} \times {\rm acceleration}$,
whereas the two terms in the RHS, $-\e^{\nu}\delta p$ and
$-\mu^l(\rho+p)\e^{\nu} H_0 Y_{lm}(\theta,\varphi)\e^{\i\omega
t}/2 $,
 correspond to the pressure  and gravity forces, respectively.
 Thus, (\ref{Lindblom-5-New}) can be understood as the
 extension of application of Newton's second law to general relativistic
 theory of stellar pulsation. Following directly from Fig.~2 and
 (\ref{Lindblom-5-New}), the net driving forces for $w$-mode
 pulsation is the sum of gravity and the Eulerian change in pressure.

 To
 compare $w$-mode with $f$ and $p$ modes, we show in Fig.~3 and Fig.~4 the corresponding
 physical quantities appearing in (\ref{Lindblom-5}) for the $f$-mode
 and the first $p$-mode of the same star. It is
 interesting to note that for $f$-mode oscillations the major
 driving force is the Eulerian change in pressure $ \delta p$ and the gravity
 in fact partially cancels the effect of $\delta p$. On the other
 hand, the major driving force for $p$-mode oscillations is
 evidently due to the Lagrangian change in pressure $ \Delta p$. We
 expect that these findings can cast light on eigen-mode analysis
 for pulsations of relativistic stars and assist in developing appropriate
 approximation schemes under different circumstances. Relevant study is now
 underway and will be reported elsewhere in due course.

\section*{Acknowledgments}
We thank LK~Tsui and LM~Lin for discussions. Our work is supported
in part by the Hong Kong Research Grants Council (Grant No: 401905
and 401807) and a direct grant (Project ID: 2060260) from the
Chinese University of Hong Kong.

\appendix

\clearpage


  \newcommand{\noopsort}[1]{} \newcommand{\printfirst}[2]{#1}
  \newcommand{\singleletter}[1]{#1} \newcommand{\switchargs}[2]{#2#1}

\clearpage
\begin{table}
\caption{Polar $w$-mode QNM frequencies of the two typical stars
considered in Fig.~1 are obtained from   GICA and exact LD
formalism, denoted respectively by $\tilde{\omega}$ and $\omega$,
and tabulated here. The percentage errors in the frequency and
damping rate, $E_{\r}=|\tilde{\omega}_\r/{\omega_\r}-1| \times 100
\%$ and $E_{\i}=|\tilde{\omega}_\i/{\omega_\i}-1| \times 100 \%$,
are also included.}
\begin{tabular}{cccccc}
\multicolumn{6}{l}{\rule[-3mm]{0mm}{8mm}$\textrm{Polytropic star}$} \\
\hline\hline \rule[-4mm]{0mm}{10mm}$M\tilde{\omega}_\r$ &
$M\tilde{\omega}_\i$ & $M\omega_\r$ & $M\omega_\i$ & $E_{\r}$ &
$E_{\i}$ \\
\hline\rule[4mm]{0mm}{0mm}
0.0360 & 0.7045 & 0.0385 & 0.7064 & 6.5 & 0.27 \\
0.3377 & 0.3804 & 0.3427 & 0.3781 & 1.5 & 0.6 \\
0.5027 & 0.2593 & 0.5029 & 0.2645 & 0.022 & 2.0 \\
0.8741 & 0.3637 & 0.8752 & 0.3694 & 0.12 & 1.5 \\
1.2235 & 0.4193 & 1.2241 & 0.4250 & 0.05 & 1.3 \\
1.5683 & 0.4596 & 1.5687 & 0.4652 & 0.026 & 1.2 \\
1.9111 & 0.4915 & 1.9114 & 0.4971 & 0.016 & 1.1 \\
2.2528 & 0.5180 & 2.2531 & 0.5235 & 0.011 & 1.1 \\
2.5938 & 0.5407 & 2.5940 & 0.5462 & 0.0077 & 1.0 \\
2.9343 & 0.5606 & 2.9344 & 0.5660 & 0.0059 & 0.97 \\
3.2743 & 0.5783 & 3.2745 & 0.5838 & 0.0051 & 0.94 \\
3.6141 & 0.5943 & 3.6143 & 0.5998 & 0.0046 & 0.91 \\
3.9536 & 0.6090 & 3.9538 & 0.6144 & 0.0041 & 0.88 \\
\hline
 &  &  &  &  &  \\
\multicolumn{6}{l}{\rule[-3mm]{0mm}{8mm}$\textrm{APR1 EOS}$} \\
\hline\hline \rule[-4mm]{0mm}{10mm} $M\tilde{\omega}_\r$ &
$M\tilde{\omega}_\i$ & $M\omega_\r$ & $M\omega_\i$ & $E_{\r}$ &
$E_{\i}$ \\
\hline\rule[4mm]{0mm}{0mm}
0.2963 & 0.3551 & 0.2983 & 0.3533 & 0.69 & 0.51 \\
0.5057 & 0.2924 & 0.5073 & 0.2963 & 0.3 & 1.3 \\
0.9247 & 0.3777 & 0.9266 & 0.3818 & 0.21 & 1.1 \\
1.3138 & 0.4307 & 1.3155 & 0.4353 & 0.13 & 1.1 \\
1.6958 & 0.4700 & 1.6973 & 0.4749 & 0.086 & 1.0 \\
2.0754 & 0.5001 & 2.0767 & 0.5051 & 0.063 & 1.0 \\
2.4551 & 0.5238 & 2.4564 & 0.5287 & 0.054 & 0.93 \\
2.8361 & 0.5431 & 2.8376 & 0.5478 & 0.053 & 0.85 \\
3.2188 & 0.5594 & 3.2205 & 0.5637 & 0.055 & 0.77 \\
3.6028 & 0.5733 & 3.6049 & 0.5771 & 0.058 & 0.66 \\
\hline
\end{tabular}
\end{table}
\clearpage
\begin{figure}
\centering
\includegraphics[width=14cm]{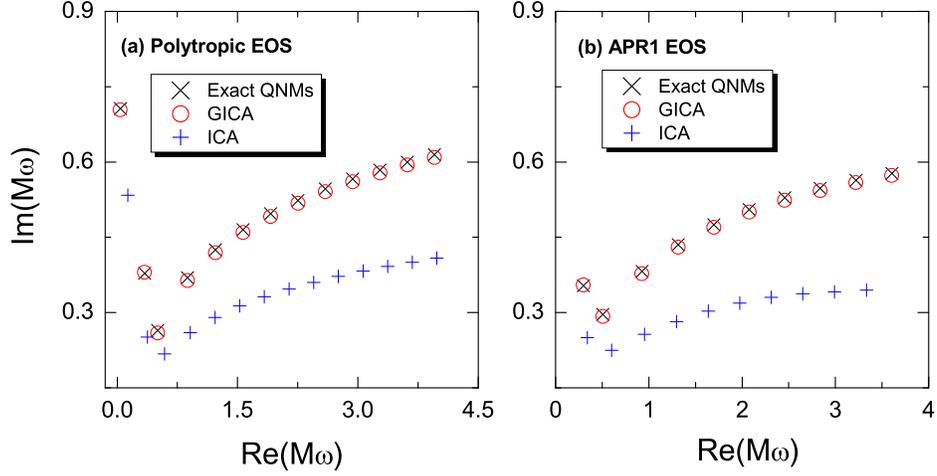}
\caption{Polar $w$-mode QNMs of (a) a polytropic star with $\gamma
=2$ and ${\rm compactness} =0.211$; and (b) an APR1 star with
${\rm compactness} =0.2$ are obtained from exact LD formalism, ICA
proposed previously and GICA developed here, respectively. Here
$M$ is the mass of the star.} \label{}
\end{figure}
\begin{figure}
\centering
\includegraphics[width=14cm]{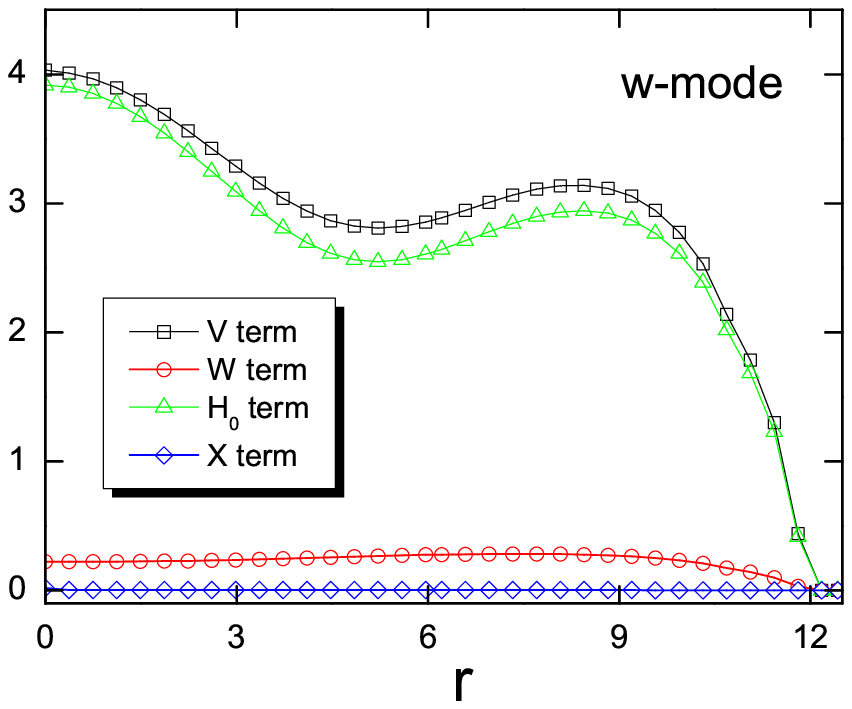}
\caption{ The absolute values of various terms in
(\ref{Lindblom-5}), $\omega^2(\rho+p)\e^{-\nu/2}V$,
$p'\e^{(\nu-\lambda)/2}W/r$, $(\rho+p)\e^{\nu/2}H_0/2$ and $X$,
are compared for the least damped polar $w$-mode ($l=2$) of the
APR1 neutron star considered in Fig.~1b.  } \label{}
\end{figure}

\begin{figure}
\centering
\includegraphics[width=14cm]{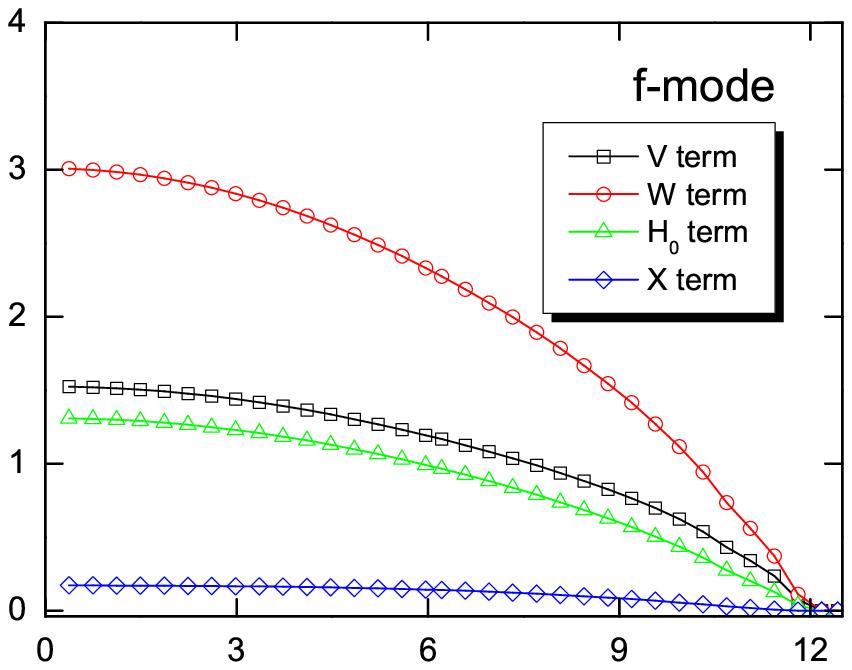}
\caption{The absolute values of various terms in
(\ref{Lindblom-5}), $\omega^2(\rho+p)\e^{-\nu/2}V$,
$p'\e^{(\nu-\lambda)/2}W/r$, $(\rho+p)\e^{\nu/2}H_0/2$ and $X$,
are compared for the $f$-mode ($l=2$) of the APR1 neutron star
considered in Fig.~1b.} \label{}
\end{figure}

\begin{figure}
\centering
\includegraphics[width=14cm]{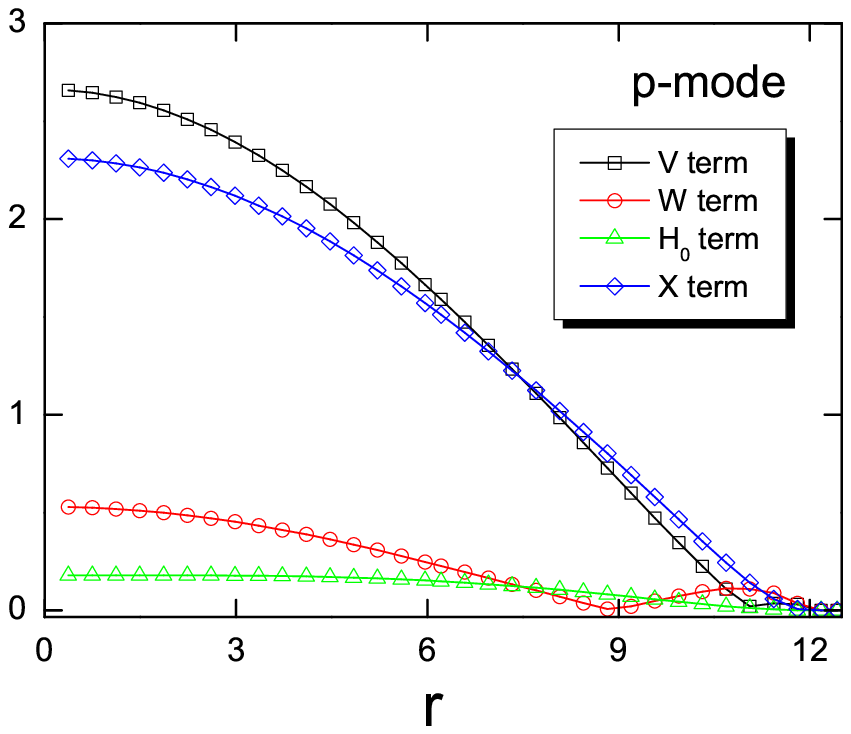}
\caption{The absolute values of various terms in
(\ref{Lindblom-5}), $\omega^2(\rho+p)\e^{-\nu/2}V$,
$p'\e^{(\nu-\lambda)/2}W/r$, $(\rho+p)\e^{\nu/2}H_0/2$ and $X$,
are compared for the first $p$-mode ($l=2$) of the APR1 neutron
star considered in Fig.~1b.} \label{}
\end{figure}

\end{document}